# *In-situ* characterization of ultrathin nickel silicides using 3D medium-energy ion scattering

Tuan Thien Tran [a], Lukas Jablonka [b], Christian Lavoie [c], Zhen Zhang [b], and Daniel Primetzhofer [a]

[a] *Department of Physics and Astronomy, Ångström Laboratory, Uppsala University, Box 516, SE-751 20 Uppsala, Sweden*

[b] *Solid State Electronics, The Ångström Laboratory, Uppsala University, SE-75121 Uppsala, Sweden*

[c] *IBM Thomas J. Watson Research Center, Yorktown Heights, New York 10598, USA*

Corresponding author: Tuan Thien Tran (tuan.tran@physics.uu.se)

## Abstract

We demonstrate a novel approach for non-destructive *in-situ* characterization of phase transitions of ultrathin nickel silicide films using 3D medium-energy ion scattering. The technique provides simultaneously composition and real-space crystallography of silicide films during the annealing process using a single sample. We show, for 10 nm Ni films on Si, that their composition follows the normal transition sequence, such as Ni-$Ni_2$Si-NiSi. For samples with initial Ni thickness of 3 nm, depth-resolved crystallography using a position-sensitive detector, shows that the Ni film transform from an as-deposited disordered layer to epitaxial silicide layers at a relatively low temperature of ~290 °C.

**Impact statement:** We demonstrate unique capabilities of 3D medium-energy ion scattering in characterization of ultrathin films. Depth-resolved composition and real-space crystallography are achieved simultaneously using a single sample and a non-invasive probe.

Metal silicides have been decisive compounds for the development of high-performance electronic devices. These materials enable low contact resistance between the devices and the interconnects, hence providing manifold benefits, such as higher drive current, lower power consumption and improved reliability. Nickel monosilicide (NiSi) has proven to be a contact material of high technological interest [1] and has been used in the 45 nm and the 32 nm transistors because it features one of the lowest resistivity of all silicides (10.5 $\mu\Omega$ cm), immunity to the fine-line effect, low formation temperature and low Si consumption [2].

It is known that the transition route of the Ni-Si system is from $Ni_2Si$ to NiSi and subsequently to $NiSi_2$ at the temperature of 800 °C [3]. However, the current extremely scaled transistor generations demand silicide film with thickness < 10 nm [1]. In this sub-10 nm regime, the full understanding of the silicide phase transitions has not yet been achieved. For example, there is a critical thickness of the initial Ni films below which the transition route is completely altered. This phenomenon was first reported in the studies of Tung *et al.* [4], in which the Ni films of 1 − 30 Å were found transformed to epitaxial layers on Si(111) and Si(100) substrates after annealing to 450 °C. Surface-sensitive methods, such as Auger electron spectroscopy and low energy electron diffraction, indicated the epitaxial phase to be $NiSi_2$. Subsequent studies using transmission electron microscopy suggested that the epitaxial films is non-stoichiometric, in which the Ni/Si ratio is about 1/1.5− 1.7 [5]. Another study suggested that the epitaxial layer might consist of small domains (a few nanometer) of both $NiSi_2$ and hexagonal θ-nickel silicide with variable Si composition of 33 − 41% [6].

As for the characterization of the materials, the commonly employed techniques are pole figure measurements (PFM) [6,7], transmission electron microscopy (TEM) [8,9] and Rutherford backscattering spectrometry (RBS) [10,11]. For example, in the PFM one acquires the diffraction response from the crystallites, hence providing the types of phases and the textural information. As the penetration depth of the X-ray used in the PFM is several microns, depth information of the silicide films is unavailable. TEM and RBS can provide in principle the depth-resolved composition and crystallography. However, as the thickness of silicide films is continuously decreasing, *in-situ* real-time characterization with high depth resolution of the films is becoming challenging for all of the above methods. This problem is one of the contributing factors for the spread of reported data of the ultrathin films. As of now, several basic questions remain open, such as: what is the composition of the epitaxial silicides and at which temperature does the epitaxial phase start to form?

As compared to the conventional ion scattering operated at MeV primary energies, Medium-Energy Ion Scattering (MEIS) can provide the depth resolution of a few nanometers [12-14]. At medium energies, detection of particles with accurate energy discrimination is possible by systems different from semiconductor detectors. Examples are electrostatic, magnetic spectrometers and multi-channel-plate delay line detectors (MCP-DLD). In combination with a pulsed beam, the latter can provide a large area position-sensitive detector permitting to acquire the X-Y positions and the energies of every particle hit (hence, the term 3D). Therefore, the blocking patterns, i.e. angular-dependent yield of the backscattered particles, in a selected range of energy can be acquired and permits to perform depth-resolved crystallography in real space [15]. Finally, as an ion scattering method, MEIS is considered non-destructive, facilitating *in-situ* characterization without affecting the results during the sequential annealing.

In this report, we demonstrate for the first time an *in-situ* study simultaneously providing the depth-resolved composition and the real-space crystallographic information of ultrathin nickel silicide films ($< 10$ nm) using 3D-MEIS. Two sets of samples with two different initial thicknesses were used: 10 nm and 3 nm Ni films on Si(100) substrates. At first, the Si substrates were cleaned using 0.5% HF solution to remove the native oxide and then immediately loaded into a magnetron sputtering tool (von Ardenne CS730S). The substrates were deposited with the Ni films using the pulse direct current magnetron sputtering: 150 W, $6 \times 10^{-3}$ mTorr Ar atmosphere and deposition rate 0.5 nm/s. After the deposition, accurate thickness measurements using RBS showed the Ni thickness to be 10.3 nm and 2.7 nm, respectively (assuming bulk density). The *in-situ* study was done with a 3D-MEIS system at the Uppsala University [15]. This system is equipped with a MCP-DLD detector (DLD120) from RoentDek. The detector is ~120 mm in diameter and located at a variable circular position 290 mm away from the samples, forming a circular acceptance angle of 21.9°. The scattering chamber is equipped with a 6-axis goniometer with 3 translational and 3 rotational degrees of freedom, orthogonal on each other, respectively. The heating filament is located underneath the samples, whereas a k-type thermocouple is placed in contact with the sample surface. When evaluating the backscattering spectra, we use SIMNRA [16] for the simulation. As multiple scattering becomes more prominent in the keV energy regime, we verified the accuracy of the SIMNRA simulation with the TRIM for backscattering (TRBS) code. TRBS is a Monte-Carlo simulation that accurately takes multiple scattering into account [16]. Both simulations are found in excellent agreement at the employed energies for the investigated Ni-Si system.

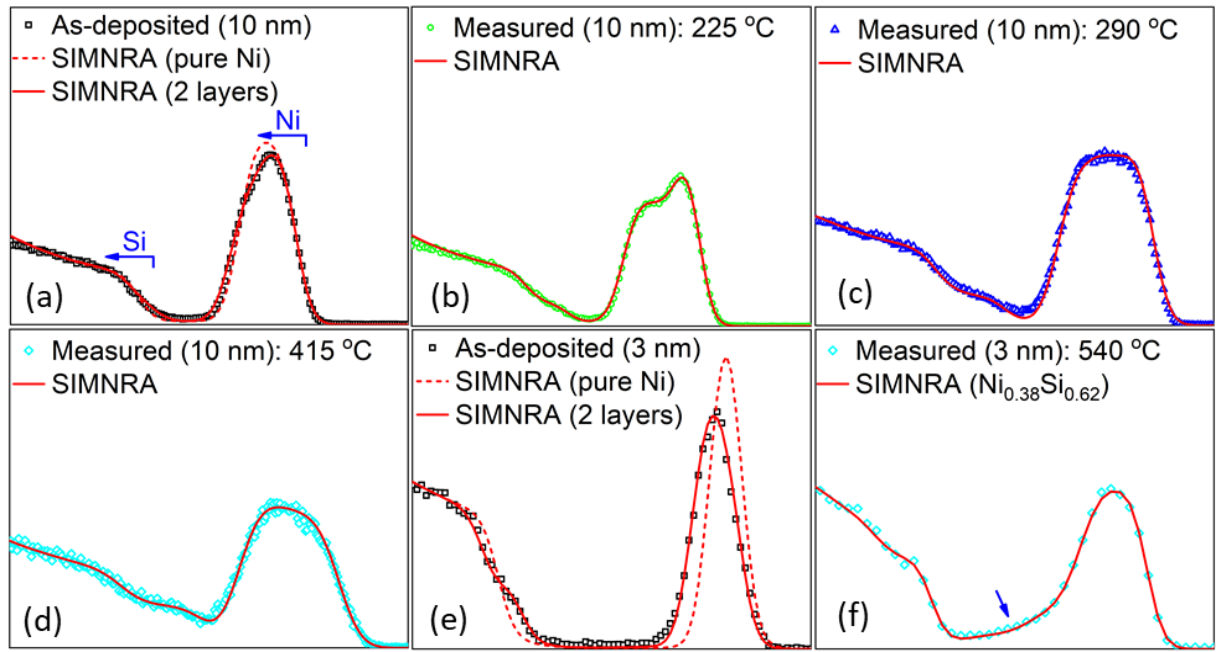

Fig. 1: Energy spectra of He backscattered from 10 nm (a-d) and 3 nm Ni on Si (100) samples (e,f) annealed at increasing temperatures. Data points represent measurements and the solid lines are simulations using SIMNRA. For details, see text.

Fig. 1 presents the spectra obtained for the 10 nm (a-d) and the 3 nm Ni films (e,f) annealed at increasing temperatures. The spectra were recorded for helium primary ions at 100 keV (a,b), 125 keV (c), 200 keV (d) and 50 keV (e,f). The energies were chosen to optimise the depth resolution while still be able to resolve the thickness of the whole films. For the composition measurement, the incidence and exit angles of the beam was set to avoid any crystal axes and planes of the crystal (complete random geometry), which is a necessary condition for the employed simulation codes. Experimental data are presented as open circles, whereas dash and solid lines are the corresponding simulation using SIMNRA. The electronic stopping power of the nickel, silicon and nickel silicide for the simulation is taken from our recent evaluation of the stopping power in the Ni-Si material system [17]. Fig. 1(a) is the spectrum of the as-deposited 10 nm Ni film. As compared to the simulation of a homogenous pure film (red-dash line), the measured spectrum features a cut-off at the top and a significant skewness of the low-energy edge of the Ni peak. This observation indicates an intermixing layer between Ni and Si at the interface. Accordingly, a two-layer model (red-solid line), in which the top layer is pure Ni and the second layer consists of 68% Ni, provides a much better fit to the measurement. The thickness of this intermixing layer is roughly half of the film. The formation of an intermixing layer between the deposited metals, such as gold [18], platinum [19] and Ni [20], is a well-known phenomenon in both sputtering and electron beam evaporation process.

In Fig. 1(b), the necessity for a two-layer model to be used in the SIMNRA simulation becomes even more apparent. The sample, for which this spectrum was recorded, was annealed to 225 °C at a rate of 25 °C/min and then kept on hold for 1 min. In the simulation shown, the top layer is of pure Ni and ~4 nm thick. Whereas, the Ni:Si ratio of the second layer is close to 2:1, in agreement with previous reports that the $Ni_2Si$ is the first phase of the Ni-Si transitions in the samples having bulk characteristics [2]. At an annealing temperature of 290 °C, the Ni peak becomes more uniform (Fig. 1(c)). Employing SIMNRA to fit this spectrum shows that the silicide phase is dominantly $Ni_2Si$. At 415 °C the silicide phase is determined to be a homogenous NiSi layer (Fig. 1(d)). For Fig. 1(e), i.e. the spectrum recorded for the as-deposited 3 nm film of Ni, a model assuming a layer of pure Ni (red-dash) cannot fully reproduce the measured data. As for the 10 nm sample, a two-layer model (red-solid), which includes an intermixing layer at the interface, provides a more accurate fit to the measurement. Finally, Fig. 1(f) shows the spectrum of the 3 nm sample annealed at 540 °C. The simulated spectrum using SIMNRA shows that a uniform silicide layer has formed with a composition of ~38% Ni and ~62% Si. There is also a subtle amount of Ni (~4%) diffusing deeper into the Si substrate (blue arrow).

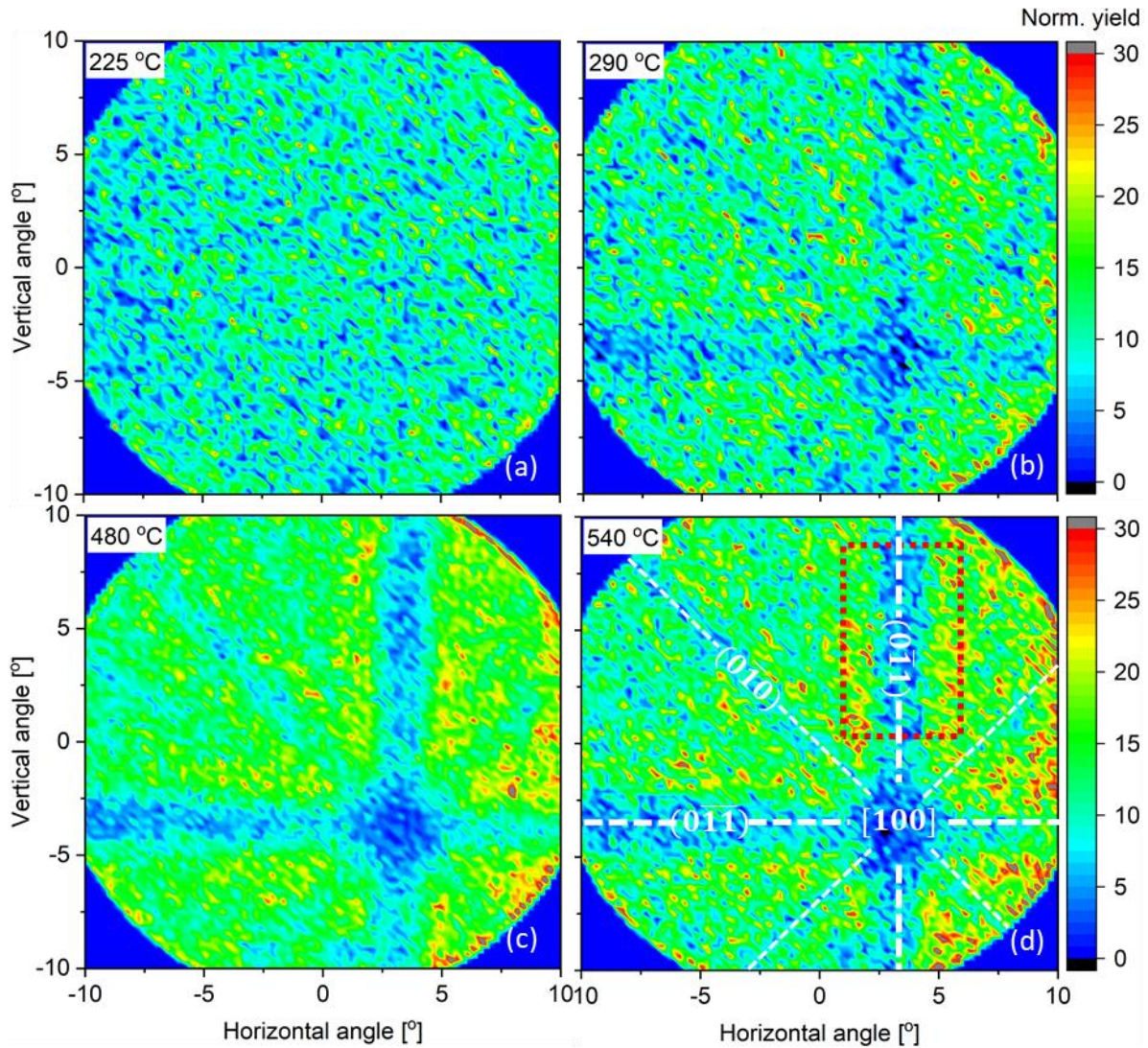

Fig. 2: Blocking patterns of the ions scattered from the 3 nm Ni on Si(100) samples annealed at increasing temperatures. The patterns shown origin from ions with scattered energies of 70 − 85 keV, equivalent to ions scattered from Ni atoms.

To obtain a more detailed picture of the transition of the 3 nm Ni films we acquired blocking patterns of the scattered ions on the detector employing the He ion beam at 100 keV primary energy. For that purpose, the sample normal was aligned close to the detector normal and the beam was incident at 40° off the sample normal. Fig. 2 shows the blocking patterns of the scattered ions within an energy window of 70 − 85 keV, corresponding to scattering from the Ni atoms. Fig. 2(a) shows the pattern upon annealing at 225 °C for 1 min. The scattered ions are distributed uniformly on the detector, indicating a macroscopically disordered atomic arrangement in the Ni-Si layer. At a temperature of 290 °C, a blocking pattern starts to appear. The definition of this pattern keeps improving at the annealing temperature of 480 °C and 540 °C. Although not shown, other blocking patterns for the energy window of the Si signal show exactly the same pattern at all temperatures. This result unequivocally demonstrates that the sputtered Ni layer was initially disordered, supposedly polycrystalline or amorphous, up to 225 ℃. The film crystallizes into an epitaxial silicide layer on the Si substrates, starting from 290 ℃.

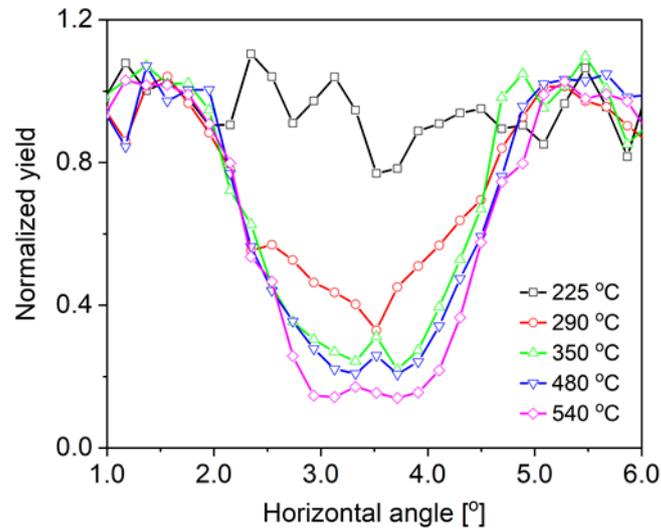

Fig. 3: Scattered yields as a function of the horizontal angle as integrated within the box in the Fig. 2(d).

To allow for a quantitative assessment of the crystallinity, Fig. 3 shows the integrated yields within the red-dot box in Fig. 2(d). The dip in the scattering yield at 290 ℃ is found shallower as compared to the one at 540 ℃. This observation can be either associated with the overall crystal quality of the film to be quite poor or the crystallization at low temperatures is starting only from the interface at 290 ℃, while the overlaying layer is still disordered. The latter case is expected as it is found that low temperature reaction between Ni and Si can form a few epitaxial monolayers at the interface while the overlayer is still disordered [21]. At 350 ℃, the crystallization is significantly more apparent and keeps improving at elevated temperature

The temperature-dependent results of the sample crystallography can be compared to earlier studies using different methods. Using AES, Tung *et al.* found that the epitaxial layer after the thermal annealing of $1 - 30$ Å Ni films on Si to be $NiSi_2$ [4]. The formation temperature of this layer was reported to be 450 ℃. Other studies using *in-situ* XRD pole figure measurement suggested that the epitaxial layer formed at 500 ℃ [22]. These temperatures are considerably higher than in our study. However, as AES is a surface-sensitive technique, it can detect the $NiSi_2$ phase only when the whole layer is transformed. Whereas, in the pole figure measurement the epitaxial layer is so thin that it is deemed difficult to detect the diffraction signal and hence the determination of the accurate temperature for the $NiSi_2$ phase might be compromised. These results show the advantages of MEIS in the study of depth-resolved real-space crystallography of the ultra-thin films because ion beam channelling is very sensitive to the near-surface crystal arrangement.

In summary, we have demonstrated the employment of the 3D-MEIS in studying *in-situ* the phase transition of ultrathin nickel silicides ($t < 10\ nm$) during thermal annealing. The advantages of the approach are the depth-resolved composition and real-space crystallography in the nanometer regime in combination with a non-invasive probe. Using the position-sensitive detector, we can, for a single sample, *in-situ* and stepwise record the transition of the initially disordered Ni film to an epitaxial layer on the Si substrate. Prior to its full transition, the epitaxial layer is expected to form only at the interface at 290 ℃. The silicide layer is fully epitaxial at the temperature of 540 ℃ and has the composition of 38% Ni and 62% Si.

**Acknowledgements:**

Support by VR-RFI (contracts #821-2012-5144 & #2017-00646_9) and the Swedish Foundation for Strategic Research (SSF, contract RIF14-0053 and SE13-0333) supporting accelerator operation is gratefully acknowledged. Zhen Zhang acknowledges the grant from the Swedish Foundation for Strategic Research (SSF, contract No. SE13-0333).